\documentclass[a4paper,oneside,11pt]{article}

\usepackage{amsfonts,color}

\newtheorem{dfn}{Definition}[section]
\newtheorem{tw}[dfn]{Theorem}
\newtheorem{prop}[dfn]{Proposition}
\newtheorem{rem}[dfn]{Remark}
\newtheorem{ex}[dfn]{Example}
\newtheorem{lem}[dfn]{Lemma}

\usepackage{amssymb} 
\usepackage{amsmath}
\numberwithin{equation}{section}

\usepackage{anysize}
\usepackage{array}
\usepackage{enumerate}

\author{Micha\l \ Barski \thanks{The paper was supported by The Polish MNiSW
grant $N N201 419039$. The second author was supported by the
European Transfer of Knowledge project SPADE2.} \\ \small  Faculty of Mathematics, Cardinal Stefan Wyszy\'nski University in Warsaw, Poland\\
\small Faculty of Mathematics and Computer Science, University of Leipzig, Germany\\ \small{\it Michal.Barski@math.uni-leipzig.de} \bigskip \\
\\
Jerzy Zabczyk
\\ \small Institute of Mathematics, Polish Academy of
Sciences,
     Warsaw,  Poland\\ \small{\it zabczyk@impan.pl}}

\title{\bf Forward rate models with linear volatilities}

\begin{document}

\maketitle

\begin{abstract}
Existence of solutions to the Heath-Jarrow-Morton equation of the
bond market with linear volatility and general L\'evy random factor
is studied. Conditions for existence and non-existence of solutions
in the class  of  bounded fields are presented. For the existence of
solutions the L\'evy process should necessarily be without the
Gaussian part and without negative jumps. If this is the case then
necessary and sufficient conditions for the existence are formulated
either in terms of the behavior of the L\'evy measure of the noise
near the origin or  the behavior of the Laplace exponent of the
noise at infinity.
\end{abstract}

\noindent
\begin{quote}


\end{quote}

\bigskip
\section{Introduction}\label{Introduction}
We are concerned with the bond market model, on a fixed time
interval $[0, T^*]$, $T^\ast<\infty$, in which the bond prices $P(t,
T),\,\,0\leq t \leq T \leq T^*\,,$ are defined on a filtered
probability space
$(\Omega,\mathcal{F},(\mathcal{F}_t),t\in[0,T^\ast],P)$ and
represented in the form
\begin{gather*}
P(t,T)= e^{-\int_{t}^{T}f(t,u)du}, \qquad t\leq T\leq T^\ast.
\end{gather*}
Thus $P(t,T)$ is the price at moment $t$ of the bond which matures
at moment $T$ and pays $1$ to the owner. The forward curves
processes $f(t,T),\,\,0\leq t \leq T \leq T^*\,,$ are It\^o
processes with stochastic differentials
\begin{gather}\label{rownanie na f}
df(t,T)=\alpha(t,T)dt+\sigma(t,T)dL(t),\qquad (t,T)\in\mathcal{T},
\end{gather}
where
\begin{gather}\label{dziedzina f}
\mathcal{T}:=\left\{(t,T)\in\mathbb{R}^2: 0\leq t\leq T\leq
T^\ast\right\},
\end{gather}
and the random factor $L$ is a real L\'evy process. This way of bond
prices modelling with $L$ replaced by a Wiener process was first
introduced by Heath, Jarrow and Morton in \cite{HJM}. The discounted
bond price process is defined by
\begin{gather*}
\hat{P}(t,T):=e^{-\int_{0}^{t}r(s)ds}\cdot P(t,T),\qquad 0\leq t\leq
T\leq T^\ast,
\end{gather*}
where $r(t):=f(t,t), \ 0\leq t\leq T^\ast$ is the short rate.
Consequently, under the assumption that $f(t,T)=f(T,T)$ for $0\leq
T<t\leq T^\ast$, we obtain the formula
\begin{gather*}
\hat{P}(t,T)=e^{-\int_{0}^{T}f(t,u)du}, \qquad 0\leq t\leq T\leq
T^\ast.
\end{gather*}
For pricing purposes it is convenient, and we do this, to require
that the discounted bond price processes $\hat{P}(\cdot,T)$ are
local martingales on $[0,T]$.

It is of  prime interest  to {\it characterize those models for
which volatility processes $\sigma$ are proportional to forward
processes}
 $f$, i.e.
\begin{gather}\label{linearity}
\sigma (t, T) =\lambda (t,T)  f(t-,T), \qquad (t,T) \in \mathcal{T},
\end{gather}
where $\lambda$ is a deterministic, positive and continuous function
on $\mathcal{T}$. As, for each $T$, $f(\cdot,T)$ is meant to be a
c\`adl\`ag process then $\sigma(\cdot,T)$ is predictable.

This problem, with $\lambda(t,T)\equiv1$, has been first stated in
\cite{Morton} in the case when $L$ was a Wiener process and solved
with a negative answer: linearity of volatility implies that there
is no  forward rate model for which \eqref{rownanie na
f}-\eqref{linearity} hold and  the discounted bond price processes
$\hat{P}(\cdot,T), \ T\in[0,T^\ast]$ are local martingales, see
\cite{Morton} Section 4.7 or \cite{Filipovic} , Section 7.4. This
fact was one of the main reasons that the BGM model was formulated
in terms of Libor rates and not in terms of forward curves, see
\cite{BGM}.

Let us recall, see  \cite{Bertoin}, \cite{Sato},
\cite{PeszatZabczyk}, that the law of the process $L$ is determined
by its  Laplace transform
\begin{gather*}\label{J}
\mathbf{E}(e^{-zL(t)})=e^{tJ(z)},\qquad t\in[0,T^\ast], \ z\in
\mathbb{R},
\end{gather*}
where
\begin{gather}\label{Laplace transform}
J(z)=-az+\frac{1}{2}qz^2+\int_{\mathbb{R}}(e^{-zy}-1+zy\mathbf{1}_{(-1,
1)}(y)) \ \nu(dy),\qquad z\in\mathbb{R},
\end{gather}
with $ a\in \mathbb{R}$, $q\geq 0$. The so called  {\it L\'evy
measure} $\nu$ satisfies integrability condition
\begin{equation}\label{war na miare Levyego}
\int_{\mathbb{R}}(y^2\wedge \ 1) \ \nu(dy)<\infty,
\end{equation}
and  $J(z)$ is a finite number if and only if  $\int_{|y| \geq
1}(e^{-zy}) \ \nu(dy)<\infty .$ The function $J$ is called the {\it
Laplace exponent} of $L$. The assumption that $\hat{P}(\cdot,T), \
T\in[0,T^\ast]$ are local martingales on $[0,T]$ implies that for
each $T\in[0,T^\ast]$, see \cite{bjork}, \cite{Eberlein},
\cite{JakubowskiZabczyk},
\begin{gather}\label{warunek HJM}
\int_{t}^{T}\alpha(t,u)du=J\left(\int_{t}^{T}\sigma(t,u)du\right)
\end{gather}
for almost all $t\in[0,T]$. So differentiating the identity
\eqref{warunek HJM} with respect to $T$, and taking into account the
condition (\ref{linearity}) we see that proportionality of the
volatility implies that the forward curve satisfies the following
equation on $\mathcal{T}$,
\begin{gather}\label{f rownanie}
df(t,T)=J^\prime \left(\int_{t}^{T}\lambda (t,u)
f(t-,u)du\right)\lambda (t,T) f(t-,T) dt+ \lambda(t,T) f(t-,T)dL(t),
\end{gather}
with the initial condition
\begin{gather}\label{initial}
f(0,T) = f_{0} (T),\,\,\,\,T \in [0, T^{*}].
\end{gather}

The paper is concerned with existence of solutions to \eqref{f
rownanie} - \eqref{initial}. We search for a solution in the class
of random fields $f(t,T), \ (t,T)\in\mathcal{T}$ such that
\begin{gather}\label{1war na rozw}
f(\cdot,T) \ \text{is adapted and
c\`adl\`ag on} \ [0,T] \ \text{for all} \ T\in[0,T^\ast],\\[1ex]\label{2war na rozw}
f(t,\cdot) \,\, \text{is continuous on} \ [t,T^\ast] \ \text{for
all} \ t\in[0,T^\ast],\\[1ex]\label{3war na rozw}
P (\sup_{(t,T)\in\mathcal{T}}f( t,T)<\infty) = 1 .
\end{gather}
Random fields satisfying \eqref{1war na rozw}-\eqref{3war na rozw}
will be called  {\it bounded fields} on $\mathcal{T}$. We also
examine a blow-up condition
\begin{gather*}
\lim_{(t,T)\rightarrow(x,y)}f(t,T)=+\infty,\qquad
(x,y)\in\mathcal{T},
\end{gather*}
in the class of locally bounded fields. For $(x,y) \in \mathcal{T}$
define
\begin{align}\label{def zbioru T_x,y}
\mathcal{T}_{x,y}:=\left\{(t,T)\in\mathcal{T}:0\leq t\leq x,0\leq
T\leq y \right\}.
\end{align}
A random field $f$ is called  {\it bounded locally on}
$\mathcal{T}_{x,y} $ if it is bounded on $\mathcal{T}_{x,y-\delta}$
for each $0<\delta<y$. \\

The results providing conditions for existence of solutions to
\eqref{f rownanie} are of two types involving either the behavior of
the function $J^\prime$ at infinity, see Theorems \ref{tw 2 glowne},
\ref{tw 1 glowne}, \ref{wniosek z tw1 o eksplozjach } or
characteristics of the noise $L$, see Theorems \ref{tw o warunkach
koniecznych}, \ref{tw glowne Tauber} and Proposition \ref{tw o
subordynatorze}. In the first case if $J^\prime$ grows slower at
infinity than a logarithm, see formula \eqref{ujemne Jprim}, then
solution exists and if $J^\prime$ grows faster than the third power
of a logarithm, see \eqref{ogr na Jprim}, then there is no solution.
The method of establishing the non-existence result - Theorem
\ref{tw 1 glowne}, is based on the approach of Morton in
\cite{Morton} where the solution is compared with a deterministic
blowing-up minorant. The paper \cite{Morton} is sketchy and the
minorant function in \cite{Morton} does not satisfy all the
conditions required in the proof. Therefore we provide a detailed
exposition with a sequence of new auxiliary results. Let us also
stress that our Theorem \ref{tw 1 glowne} treats the problem for a
general class of functions $J^\prime$. In the special case of
bounded $J^\prime(z), z\geq 0$ the existence result given by Theorem
\ref{tw 2 glowne} can be deduced, via Musiela parametrization, from
the results presented in \cite{PeszatZabczyk}. The second group of
results, involving characteristics of the noise $L$, is deduced from
Theorem \ref{tw 2 glowne}  and \ref{tw 1 glowne}. It turns out that
if the equation \eqref{f rownanie} has a solution then necessarily
the process $L$ does not have a Gaussian part and its jumps must be
positive, see Theorem \ref{tw o warunkach koniecznych}. If this is
the case then necessary and sufficient conditions for existence are
formulated in terms of the behavior of the function
$$
U_{\nu}(x):=\int_{0}^{x}y^2\nu(dy),
$$
near $0$, see Theorem \ref{tw glowne Tauber}. An essential role here
is played by a Tauberian theorem, see Theorem 2 p.445 in
\cite{Feller}. An existence result for integrable subordinators is
formulated as Proposition \ref{tw o subordynatorze}.

Although there are severe restrictions on the noise required for
existence of the solution nevertheless the class of models with
linear volatilities allows to describe bond market with upward and
downward price movements and therefore might be useful for
applications.

 We approached the existence problem by working with
the theory of random fields because this way minimal requirements
are imposed on the model. It is of great interest to compare this
approach with that using stochastic partial differential equations
like in the papers \cite{FilipovicTappe},
\cite{FilipovicTappeTeichmann}, \cite{Marinelli},
\cite{peszaZabczyk-preprint}, \cite{PeszatZabczyk}. In the paper
under preparation \cite{BarskiZabczyk} the existence problem for
forward rates with linear volatilities in the weighted spaces of
square integrable functions and functions with square integrable
first derivative via the theory of random fields is examined.

The paper is organized as follows. Section \ref{Preliminaries}
contains preliminaries and reformulation of the problem to the more
tractable form. Section \ref{Formulation of the main results} is
devoted to the presentation of the main general results. Section
\ref{Corollaries and comments} contains corollaries, examples and
comments regarding larger class in which solution can be searched.
Proofs are postponed to Section \ref{Proofs}.

\vskip2mm
\noindent
{\bf Acknowledgements}\\
The authors thank Damir Filipovi\'c for providing a copy of
\cite{Morton} and a section of a book \cite{Filipovic} which was
under preparation while writing this paper. The second author thanks
Szymon Peszat for a useful discussion on the subject of the paper.
Constructive comments of reviewers and editors of F\&S on the
initial and the corrected versions of the paper are also gratefully
acknowledged.

\section{Model settings}\label{Preliminaries}
Here we introduce notation and  assumptions needed in the sequel and
transform  the equation (\ref{f rownanie}) to a form easier to
investigate.

We set the notation
\begin{gather}\label{ograniczenia lambdy}
\underline{\lambda}:=\inf_{(t,T)\in\mathcal{T}}\lambda(t,T)>0,
\qquad
\bar{\lambda}:=\sup_{(t,T)\in\mathcal{T}}\lambda(t,T)<+\infty.
\end{gather}
As we intend to work with positive forward rates we introduce the
following assumptions, compare \eqref{basic} below.
\begin{enumerate}[]
\item (A1) \quad The initial curve $f_{0}$ is positive and continuous on $[0,T^\ast]$.
\item (A2) \quad The support of the L\'evy measure is contained in the interval
    $(-1/\bar {\lambda},+\infty)$.
\end{enumerate}
\begin{prop}
Assume that $f$ is a bounded field and conditions (A1) and (A2) are
satisfied. Then $f$ is a solution of \eqref{f rownanie} if and only
if
\begin{equation}\label{basic3}
f(t,T) =
a(t,T)e^{\int_{0}^{t}J^\prime\left(\int_{s}^{T}\lambda(s,u)f(s,u)du\right)\lambda(s,T)ds},\qquad
(t,T) \in
      \mathcal{T},
\end{equation}
where, for $(t,T) \in
      \mathcal{T},$
\begin{gather}\label{wzor na a}
a(t,T):=f_{0}(T)
 e^{\int_{0}^{t} \lambda (s, T) dL(s)-\frac{q^2}{2}\int_{0}^{t}\lambda^2(s,T)ds + \int_{0}^{t}\int_{-1/{\bar {\lambda}}}^{+\infty} \big(\ln (1+\lambda (s,T)y) -\lambda (s,T)y\big)\pi (ds, dy)}.
\end{gather}
\end{prop}
{\bf Proof:} Let us notice, that for each $T$ the solution  $f(t,
T),\,\, t\in [0,T]$ of  \eqref{f rownanie} is a stochastic
exponential and therefore, see Theorem II.37 in \cite{Protter},
equation \eqref{f rownanie} can be equivalently written as
\begin{multline}\label{basic}
\hspace*{1cm}f(t,T)=f_{0}(T)\,\, e^{\int_{0}^{t}J^\prime\left(\int_{s}^{T}\lambda(s,u)f(s-,u)du\right)\lambda(s,T)ds+\int_{0}^{t} \lambda (s, T) dL(s)-\frac{q^2}{2}\int_{0}^{t}\lambda^2(s,T)ds}\\
   \hspace*{-1cm}   \cdot\prod_{s\leq
t}(1+\lambda (s,T)\triangle L(s))e^{-\lambda (s,T)\triangle
L(s)},\qquad (t,T) \in \mathcal{T},
\end{multline}
where $\triangle L(s) = L(s) - L(s-)$. Let us recall that if $L$ is
a L\'evy process then it can be decomposed into the L\'evy-It\^o
form, see \cite{Apl},
\begin{gather}\label{rownanie na szum}
L(t):=at+qW(t)+\int_{0}^{t}\int_{|y|<1} y \
\hat{\pi}(ds,dy)+\int_{0}^{t}\int_{|y|\geq 1}y \ {\pi}(ds,dy),
\end{gather}
where $a\in\mathbb{R}, q\geq0$, $W$ is a Wiener process, $\pi$ is
the Poisson random measure of jumps of $L$  and $\hat{\pi}$ is the
measure $\pi$ compensated by $dt \times \nu (dy)$.

Under assumptions (A1) and (A2) we can write equation (\ref{basic})
in the form
\begin{multline}\label{basic1}
f(t,T)=f_{0}(T)\,\,
e^{\int_{0}^{t}J^\prime\left(\int_{s}^{T}\lambda(s,u)f(s-,u)du\right)\lambda(s,T)ds+\int_{0}^{t}
\lambda (s, T)
dL(s)-\frac{q^2}{2}\int_{0}^{t}\lambda^2(s,T)ds}\\[2ex]
      \cdot e^{\int_{0}^{t}\int_{-1/{\bar {\lambda}}}^{+\infty} \big(\ln (1+\lambda (s,T)y) -\lambda (s,T)y \big)\pi (ds, dy)},\qquad (t,T) \in
      \mathcal{T},
\end{multline}
or equivalently as
\begin{equation}\label{basic2}
f(t,T) =
a(t,T)e^{\int_{0}^{t}J^\prime\left(\int_{s}^{T}\lambda(s,u)f(s-,u)du\right)\lambda(s,T)ds},\qquad
(t,T) \in
      \mathcal{T}.
\end{equation}
We show now that  we can replace $f(s-,u)$ in \eqref{basic2} by
$f(s,u)$. To do this we prove that for each $(t,T) \in \mathcal{T}$
\begin{gather*}
\int_{0}^{t}J^\prime\left(\int_{s}^{T}\lambda(s,u)f(s,u)du\right)\lambda(s,T)ds=
\int_{0}^{t}J^\prime\left(\int_{s}^{T}\lambda(s,u)f(s-,u)du\right)\lambda(s,T)ds.
\end{gather*}
Let us start with the observation that for $T\in[0,T^\ast]$ moments
of jumps of the process $f(\cdot,T)$ are the same as for
$a(\cdot,T)$. Moreover, it follows from \eqref{wzor na a} that the
set of jumps of $a(\cdot,T)$ is independent of $T$ and is contained
in the set
\begin{gather*}
\mathcal{Z}:=\{t\in[0,T^\ast]: \triangle L(t)\neq0\}.
\end{gather*}
Thus if $s\notin \mathcal{Z}$ then
\begin{gather*}
J^\prime\left(\int_{s}^{T}\lambda(s,u)f(s,u)du\right)\lambda(s,T)=
J^\prime\left(\int_{s}^{T}\lambda(s,u)f(s-,u)du\right)\lambda(s,T).
\end{gather*}
By Theorem 2.8 in \cite{Apl} the set $\mathcal{Z}$ is at most
countable, so the assertion follows.\hfill$\square$

\begin{rem}\label{positivity1}
As we already indicated, the formula \eqref{basic} implies that
models with positive forward rates must satisfy (A1) and (A2). If
this is the case then, in view of \eqref{basic3}, only properties of
the restriction of the function $J^\prime$ to $[0,+\infty)$ are
essential for the existence results.
\end{rem}

As far as the coefficient function $\lambda$ in the equation
\eqref{basic3} is concerned, we will require that it is continuous
function satisfying the following assumption.
\[(A3) \quad
\begin{cases}
\ \text{For each} \ 0<t\leq T^\ast \ \text{the process} \
\int_{0}^{t}\lambda(s,T)dL(s); \ T\in[t,T^\ast] \ \text{is
continuous.} \\[2ex]
\ \text{The field} \ \mid\int_{0}^{t}\lambda(s,T)dL(s)\mid \
\text{is bounded on} \ \mathcal{T}.
\end{cases}
\]
Compare the type of boundeddness above with \eqref{3war na rozw}.
The condition $(A3)$ is satisfied if, for instance,
$\lambda(\cdot,\cdot)$ is constant or, more generally, if it is of
the form
\begin{gather*}
\lambda(t,T)=\sum_{n=1}^{N}a_n(t)b_n(T),
\end{gather*}
where $\{a_n(\cdot)\},\{b_n(\cdot)\}$ are continuous functions. The
assumption that $\lambda(\cdot,\cdot)$ is continuous does not imply,
in general, (A3), see \cite{Brzezniak}, \cite{Kwapien} for
counterexamples.

\begin{prop}\label{prop ograniczenie a}
Assume that the conditions (A1), (A2), (A3) are satisfied. Then  the
field $\{a(t,T); \ (t,T)\in\mathcal{T}\}$ given by \eqref{wzor na a}
is bounded from below and above by strictly positive constants
depending on $\omega$. Moreover, $a(\cdot,T)$ is adapted and
c\`adl\`ag on $[0,T]$ for all $T\in[0,T^\ast]$ and $a(t,\cdot)$ is
continuous on $[t,T^\ast]$ for all $t\in[0,T^\ast]$.
\end{prop}
{\bf Proof:} The fact that $a(\cdot,T)$ is adapted and c\`adl\`ag is
clear. We only need to show that
\begin{gather*}
F(t,T,\lambda(\cdot,\cdot)):=\int_{0}^{t}\int_{-\frac{1}{\bar{\lambda}}}^{+\infty}\Big(\ln(1+\lambda(s,T)y)-\lambda(s,T)y\Big)\pi(ds,dy)
\end{gather*}
is bounded wrt. $(t,T)$ and continuous wrt. $T$. The function
$x\longrightarrow \ln(1+xy)-xy$ is decreasing on
$[\underline{\lambda},\bar{\lambda}]$ and thus we have
\begin{gather*}
F(t,T,\bar{\lambda})\leq F(t,T,\lambda(\cdot,\cdot))\leq
F(t,T,\underline{\lambda}).
\end{gather*}
The processes $F(t,T,\bar{\lambda}),F(t,T,\underline{\lambda})$ do
not depend on $T$ and have c\`adl\`ag paths in $t$. Therefore they
are bounded wrt. $t$. Continuity of $F$ wrt. $T$ follows from the
dominated convergence theorem.~\hfill$\square$

Let us focus on the function $J^\prime$ appearing in the equation
\eqref{basic3}. In virtue of \eqref{Laplace transform},
\eqref{rownanie na szum} and by the assumption (A2) the function $J$
is given by the formula
\begin{gather}\label{J u nas}\nonumber
J(z)=-az+\frac{1}{2}qz^2+J_1(z)+J_2(z)+J_3(z),
\end{gather}
where
\begin{gather}
J_1(z):=\int_{-1/{\bar {\lambda}}}^{0}(e^{-zy}-1+zy) \ \nu(dy),\quad
J_2(z):=\int_{0}^{1}(e^{-zy}-1+zy) \
\nu(dy)\\[2ex]
J_3(z):=\int_{1}^{\infty}(e^{-zy}-1) \ \nu(dy).
\end{gather}
Taking into account \eqref{war na miare Levyego} we see that the
function $J$ is well defined for  $z\geq 0$.  Moreover, the
condition \eqref{war na miare Levyego} implies that for $z>0$ the
functions $J_1,J_2,J_3$ have derivatives of any order, see Lemma 8.1
and 8.2 in \cite{Rusinek}. In the equation (\ref{f rownanie}) or
equivalently in (\ref{basic3}) intervene values of $J^\prime(z)$ for
{\it all non-negative} $z$. We will therefore assume that also
$J^\prime(0)$ is a finite number. But
\begin{gather*}
\quad J^\prime(0) =-a+J^\prime_3(0)=-a-\int_{1}^{\infty}y \nu (dy),
\end{gather*}
so we require that
\begin{equation}\label{condition on intensity1}
\int_{1}^{\infty}y \nu (dy) < +\infty.
\end{equation}

Thus the objective of this paper is to examine existence of solution
for the equation (\ref{basic3}), where
\begin{gather*}
J^\prime(z)=-a+qz+J^\prime_1(z)+J^\prime_{2}(z)+J^\prime_3(z),\qquad
z\geq 0,
\end{gather*}
and the L\'evy measure $\nu$ of $L$ is concentrated on $(-1/{\bar
{\lambda}}, 0)\cup(0,+\infty)$ and satisfies the assumption
\begin{gather*}
\text{(A4)}\qquad\qquad \int_{(-1/{\bar {\lambda}}, 1)}y^2 \nu(dy) +
\int_{1}^{\infty} y \nu (dy) <\infty.
\end{gather*}
The following properties of the function $J^\prime$ will be needed
in the sequel.
\begin{prop}\label{prop Jprim rosnaca}
\begin{enumerate}[i)]
\item If (A4) holds then $J_{1}^{\prime},J_{2}^{\prime},J_{3}^{\prime}$, and thus
${J^\prime}$ as well, are increasing, real-valued functions on the
interval $[0,+\infty)$.
\item $J^\prime$ is a Lipschitz function on $[0,+\infty)$ if and only if
\begin{equation}\label{square integrability}
\int_{1}^{\infty} y^{2} \nu (dy) <\infty.
\end{equation}
\end{enumerate}
\end{prop}
{\bf Proof:} The proof follows directly from the formulae for the
derivatives of $J_{1},J_{2},J_{3}$ listed below, see Lemma 8.1 and
8.2 in \cite{Rusinek}
\begin{gather}\label{pierwsze pochodne}\nonumber
J_1^{\prime}(z)=\int_{-1/{\bar {\lambda}}}^{0}y(1-e^{-zy})\nu(dy),
\quad J_2^{\prime}(z)=\int_{0}^{1}y(1-e^{-zy})\nu(dy),\\[2ex]
J_3^{\prime}(z)=-\int_{1}^{\infty}ye^{-zy}\nu(dy),
\end{gather}
\begin{gather}\label{drugie pochodne}
J_1^{\prime\prime}(z)=\int_{-1/{\bar
{\lambda}}}^{0}y^2e^{-zy}\nu(dy), \quad
J_2^{\prime\prime}(z)=\int_{0}^{1}y^2e^{-zy}\nu(dy), \quad
J_3^{\prime\prime}(z)=\int_{1}^{\infty}y^2e^{-zy}\nu(dy).
\end{gather}
\hfill$\square$

\begin{rem}
The conditions (A4) and (\ref{square integrability}) are equivalent
to the, respectively, integrability and square integrability of the
process $L$, see \cite{Sato}, Theorem 2.53 and Proposition 25.4
p.159.
\end{rem}

\section{Formulation of the results}\label{Formulation of the main results}

In this section we present main results providing conditions for
existence and non-existence of solution of the equation
\eqref{basic3}. The first two results, i.e. Theorem \ref{tw o
warunkach koniecznych} and Theorem \ref{tw glowne Tauber} are
formulated in terms of the characteristics of the process $L$. Both
are deduced from Theorem \ref{tw 2 glowne} and Theorem \ref{tw 1
glowne} which provide conditions for existence and non-existence of
solutions in terms of the growth of the function $J^\prime$ at
infinity. The final result is Theorem \ref{wniosek z tw1 o
eksplozjach } on locally bounded solutions which were defined in
Section \ref{Introduction}. In section \ref{Corollaries and
comments} we present a further result regarding subordinators -
Proposition \ref{tw o subordynatorze}, which is based on the results
from this section. Proofs are contained in Section \ref{Proofs}.

In the formulation of all the results we implicitly assume that
(A1), (A2), (A3), (A4) are satisfied.

The first theorem states that for existence of bounded solutions to
\eqref{basic3} the Gaussian part of the noise process $L$ must be
absent and, rather unexpectedly, $L$ must not have negative
jumps.\vspace{2mm}

\begin{tw}\label{tw o warunkach koniecznych}
If the Laplace exponent $J$ of $L$ is such that $q>0$ or
$\nu\{(-\frac{1}{\bar{\lambda}},0)\}>0$ then there are  no bounded
solutions to \eqref{basic3}.
\end{tw}

To go further we therefore assume that $q=0$ and that the support of
$\nu$ is contained in $[0, +\infty).$ It turns out that then the
solution of the problem is related to the behavior  of the
distribution function
\begin{gather*}
U_{\nu}(x):=\int_{0}^{x}y^2\nu(dy),\qquad x\geq0,
\end{gather*}
of the modified L\'evy measure $y^2\nu(dy)$ near the origin. For the
formulation we need the   concept of slowly varying functions. A
positive function $M$ {\it varies slowly at $0$} if for any fixed
$x>0$
\begin{gather*}
\frac{M(tx)}{M(t)}\longrightarrow 1, \qquad \text{as} \
t\longrightarrow0.
\end{gather*}
Typical examples are constants or, for arbitrary $\gamma$ and small
positive $t$, functions
$$
M(t) = \left(\ln {\frac {1}{t} }\right)^{\gamma}\,\,\,.
$$
If $M$ varies slowly at zero, then for any $\varepsilon>0$ the
following estimation holds, see Lemma 2 p.277 in \cite{Feller},
\begin{gather}\label{ograniczenie funkcji wolno zmieniajacej}
t^{\varepsilon}<M(t)<t^{-\varepsilon}\,\,,
\end{gather}
for all positive $t$ sufficiently small. If
\begin{gather*}
\frac{f(x)}{g(x)}\longrightarrow 1, \qquad \text{as} \
x\longrightarrow0,
\end{gather*}
then we write $f(x)\sim g(x)$.

\begin{tw}\label{tw glowne Tauber}
Assume that for some $\rho\in(0,+\infty)$,
\begin{gather}\label{warunek na zachowanie U w zerze}
U_{\nu}(x)\sim  x^{\rho} \cdot M(x), \qquad as \ x\rightarrow0,
\qquad
\end{gather}
where $M$ is a slowly varying function at $0$.
\begin{enumerate}[i)]
\item If $\rho>1$ then there exists a bounded solution of \eqref{basic3}.
\item If $\rho<1$, then there is no bounded solution of \eqref{basic3}.
\item If $\rho=1$, the measure $\nu$ has a density and
\begin{equation}\label{warunek na L}
M(x) \longrightarrow 0 \quad \text{as} \ x\rightarrow 0, \quad and
\quad \int_{0}^{1}\frac{M(x)}{x}\ dx=+\infty,
\end{equation}
then there exists a bounded solution of \eqref{basic3}.
\end{enumerate}
\end{tw}

The following two characterizations are of independent interest and
are  crucial for the proofs of all the results presented above. We
set $\mathbb{R}_{+}=[0,+\infty)$.

\begin{tw}\label{tw 2 glowne}
Assume that
\begin{gather}\label{ujemne Jprim}
 \limsup_{z\rightarrow\infty} \ \left(\ln z-\bar {\lambda} T^\ast J^\prime(z)\right)=+\infty.
\end{gather}
\begin{enumerate}[i)]
\item Then there exists a bounded field
$f:\mathcal{T}\longrightarrow\mathbb{R}_{+}$ which solves
\eqref{basic3}.
\item If, in addition, \eqref{square integrability} holds then the
solution $f$ is unique in the class of bounded fields.
\end{enumerate}
\end{tw}

\begin{tw}\label{tw 1 glowne}
Assume that for some $a>0$, $b\in\mathbb{R}$,
\begin{gather}\label{ogr na Jprim}
J^\prime(z)\geq a(\ln z)^3+b, \qquad \forall z>0.
\end{gather}
For arbitrary $\kappa \in (0,1)$, there exists a positive constant
$K$ such that if
\begin{gather}\label{oddzielenie f_0}
f_{0}(T)> K, \quad\forall T\in[0,T^\ast],
\end{gather}
then there is no solution
$f:\mathcal{T}\longrightarrow\mathbb{R}_{+}$ of the equation
\eqref{basic3} which is bounded with probability greater or equal
than $\kappa$.
\end{tw}

The proof of Theorem \ref{tw 1 glowne} implies a  result on locally
bounded solutions. Namely,  a locally bounded solution necessarily
blows up in some point of the domain $\mathcal{T}$.
\begin{tw}\label{wniosek z tw1 o eksplozjach }
Let all the assumptions of Theorem \ref{tw 1 glowne} be satisfied.
There exists a point $(x,y)\in\mathcal{T}$ such that if $f$ is a
solution of \eqref{basic3} which is bounded locally on
$\mathcal{T}_{x,y}$ then
\begin{gather*}
\lim_{\mathcal{T}_{x,y}\ni(t,T)\rightarrow(x,y)}f(t,T)=+\infty
\end{gather*} with probability
greater or equal than $\kappa$.
\end{tw}

\section{Corollaries and comments}\label{Corollaries and comments}

\subsection{Existence of solutions and subordination}
Subordinator is an increasing L\'evy process. Its L\'evy measure
$\nu$ is concentrated on a positive half-line and satisfies
condition
\begin{gather}\label{warunek na pierwszy moment miary}
\int_{0}^{1}y \nu(dy)<\infty,
\end{gather}
see \cite{Apl} Theorem 1.3.15 p.49 and \cite{Sato} Proposition 21.7
p. 137. If $L$ is subordinator then, as the proposition below
states, the equation \eqref{basic3}, has a bounded solution.
However, \eqref{warunek na pierwszy moment miary} is not necessary
for the existence as Example \ref{przyklad z subordynatorem} shows.
This way we answer a question posed by one of the reviewers.

\begin{prop}\label{tw o subordynatorze}
If the process $L$ is a sum of a subordinator and a linear function
then \eqref{basic3} has a bounded solution. In particular if $L$ is
a compound Poisson process with a drift and positive jumps only then
$\eqref{basic3}$ has a bounded solution.
\end{prop}
\noindent {\bf Proof:} By direct calculation we have
\begin{gather*}
J^\prime(z)=-a+\int_{0}^{1}y(1-e^{-zy})\nu(dy)-\int_{1}^{\infty}ye^{-zy}\nu(dy)\leq-a+\int_{0}^{1}y\nu(dy).
\end{gather*}
Thus $J^\prime$ is a bounded function, therefore satisfies
\eqref{ujemne Jprim} and the result follows from Theorem \ref{tw 2
glowne}.\hfill$\square$

\begin{ex}\label{przyklad z subordynatorem}
Let
\begin{gather*}
\nu(dy)=\frac{1}{y^2\mid\ln
y\mid^\gamma}\mathbf{1}_{(0,\frac{1}{2})}(y)dy
\end{gather*}
where $\gamma>0$. Then the following hold.
\begin{enumerate}[a)]
\item There exists a bounded solution for any $\gamma>0$.
\item $\int_{0}^{1}y \nu(dy)<\infty \Longleftrightarrow
\gamma>1$.
\end{enumerate}
\end{ex}
{\bf Proof:} We find $J_2$ explicitly. After some calculations we
obtain
\begin{align*}
J_2(z)=\int_{0}^{\frac{1}{2}}y(1-e^{-zy})\frac{1}{y^2\mid\ln
y\mid^\gamma}dy=\int_{0}^{\frac{z}{2}}\frac{1-e^{-u}}{u}\cdot\frac{1}{\mid\ln\frac{u}{z}\mid^\gamma}du.
\end{align*}
For large $z$ we have
\begin{align*}
J_2(z)\leq
c\int_{0}^{\frac{1}{2}}\frac{1}{\mid\ln\frac{z}{u}\mid^\gamma}du+\int_{\frac{1}{2}}^{\frac{z}{2}}\frac{1}{u}\cdot\frac{1}{\mid\ln\frac{z}{u}\mid^\gamma}du.
\end{align*}
The first integral tends to $0$ with $z \rightarrow +\infty$. The
second can be written in the form
\begin{gather*}
\int_{\frac{1}{2}}^{\frac{z}{2}}\frac{1}{u}\cdot\frac{1}{\mid\ln\frac{z}{u}\mid^\gamma}du=\int_{4}^{2z}\frac{1}{v
\mid\ln v\mid^\gamma}dv.
\end{gather*}
As a consequence
\begin{gather*}
\lim_{z\rightarrow +\infty}\frac{J^\prime_2(z)}{\ln z}=0
\end{gather*}
and thus
\begin{gather*}
\lim_{z\rightarrow +\infty}\left(\ln z-aJ^\prime_2(z)\right)=+\infty
\end{gather*}
for any $a>0$. Thus \eqref{ujemne Jprim} holds and solution exists.
Checking $(b)$ is straightforward.\hfill $\square$\vspace{3mm}

\subsection{Comments on Theorem \ref{tw glowne Tauber}}

We formulate two examples for which the conditions
\begin{equation}\label{pierwszy warunek na L}
M(x) \longrightarrow 0 \qquad \text{as} \ x\rightarrow 0,\,\,\,
\end{equation}
\begin{equation}\label{drugi warunek na L}
\int_{0}^{1}\frac{M(x)}{x}\ dx=+\infty,
\end{equation}
are not simultaneously satisfied but the existence problem can be
solved in virtue of Theorem \ref{tw 2 glowne}.

\begin{ex}
Let $\nu$ be a measure with density
\begin{gather*}
\nu(dx)=\frac{1}{x^2}\cdot\frac{(\ln\frac{1}{x})^\gamma+\gamma(\ln
\frac{1}{x})^{\gamma-1}}{(\ln\frac{1}{x})^{2\gamma}}\cdot\mathbf{1}_{(0,1)}(x),
\qquad \gamma>1.
\end{gather*}
Then it can be checked that the function $U_{\nu}$ is given by
\begin{gather*}
U_{\nu}(x)=\int_{0}^{x}y^2\cdot
g(y)dy=x\cdot\frac{1}{(\ln\frac{1}{x})^\gamma}.
\end{gather*}
It is clear that the function
\begin{gather*}
M(x):=\frac{1}{(\ln \frac{1}{x})^{\gamma}}, \quad \gamma>1,
\end{gather*}
varies slowly at zero and that \eqref{pierwszy warunek na L} holds.
However, condition \eqref{drugi warunek na L} is not satisfied and
thus Theorem \ref{tw glowne Tauber} does not cover this case. We can
explicitly show that $J^\prime_2$ is bounded and use Theorem \ref{tw
2 glowne}. We have
\begin{align*}
J^\prime_2(z)&=\int_{0}^{1}y(1-e^{-zy})g(y)dy=\int_{1}^{+\infty}\frac{1-e^{-\frac{z}{x}}}{x}\cdot\frac{(\ln
x)^\gamma+\gamma(\ln x)^{\gamma-1}}{(\ln x)^{2\gamma}}dx\\[2ex]
&\leq \int_{1}^{+\infty}\frac{1}{x}\cdot\frac{(\ln
x)^\gamma+\gamma(\ln x)^{\gamma-1}}{(\ln x)^{2\gamma}}dx<+\infty.
\end{align*}
\end{ex}\vspace{2mm}

\begin{ex}\label{przyklad4}
Let $\nu$ be given by
\begin{gather*}
\nu(dy)=\frac{1}{y^{1+\rho}}\mathbf{1}_{(0,1)}(y) \ dy, \qquad
\rho\in(0,2).
\end{gather*}
Then
\begin{enumerate}[a)]
\item if $\rho\in(1,2)$ then  equation \eqref{basic3} has
    no bounded solutions,
\item if $\rho\in(0,1)$  or
\item  $\rho=1$ and $\bar{\lambda}T^\ast<1$ then equation  \eqref{basic3} has a bounded
solution.
\end{enumerate}
\end{ex}
{\bf Proof:} For $\rho\in(0,2)$ we have
\begin{gather*}
U_{\nu}(x)=\frac{1}{2-\rho}x^{2-\rho}, \qquad x\in(0,1),
\end{gather*}
and thus $(a)$ and $(b)$ follows from Theorem \ref{tw glowne
Tauber}. If $\rho=1$ than $(c)$ can not be deduced from Theorem
\ref{tw glowne Tauber} because the function $M(x)\equiv 1$ does not
tend to zero. However, we have
\begin{gather*}\label{Jprim w neg przykl 4}\nonumber
J^\prime_{2}(z)=\int_{0}^{1}y(1-e^{-zy})\frac{1}{y^{2}} \ dy
=\int_{0}^{z}\frac{1-e^{-v}}{\frac{v}{z}}\frac{1}{z} \
dv=\int_{0}^{z}\frac{1-e^{-v}}{v} \ dv,
\end{gather*}
and consequently
\begin{align*}
\lim_{z\rightarrow\infty}\frac{\ln z}{\bar{\lambda}T^\ast
J^\prime_2(z)}\overset{d'H}{=}\lim_{z\rightarrow\infty}\frac{\frac{1}{z}}{\frac{1-e^{-z}}{z}\cdot{\bar{\lambda}T^\ast}}
=\lim_{z\rightarrow\infty}\frac{1}{\bar{\lambda}T^\ast(1-e^{-z})}=\frac{1}{\bar{\lambda}T^\ast}>1.
\end{align*}
This condition clearly implies $\eqref{ujemne Jprim}$ and $(c)$
follows from Theorem \ref{tw 2 glowne}. \hfill$\square$

\subsection{Integrable solutions}

In the case when there is no solution of equation \eqref{basic3} in
the class of bounded fields then one may ask if the solution does
exist in a wider class of fields satisfying some integrability
conditions. However, in some situations these two classes are the
same. Assume, for instance, that the solution is supposed to satisfy
the integrability condition
\begin{gather*}
\int_{0}^{T^\ast}J^\prime\left(\bar{\lambda}\int_{0}^{T^\ast}f(s,u)\
du\right)ds<\infty, \qquad P-a.s..
\end{gather*}
By Proposition \ref{prop Jprim rosnaca} the function $J^\prime$ is
increasing on $[0,+\infty)$, so is the nonnegative function
$J^\prime(\cdot)+\mid J^\prime(0)\mid$. Consequently, for any
$(t,T)\in\mathcal{T}$ we have
\begin{align*}
f(t,T)&=e^{\int_{0}^{t}J^\prime\left(\int_{s}^{T}f(s,u)\lambda(s,u)du\right)\lambda(s,T)ds}\cdot
a(t,T)\\[2ex]
&\leq
e^{\int_{0}^{t}\left(J^\prime\left(\int_{s}^{T}f(s,u)\lambda(s,u)du\right)+\mid
J^\prime(0)\mid\right)\lambda(s,T)ds}\cdot
a(t,T)\\[2ex]
&\leq
e^{\bar{\lambda}\int_{0}^{T^\ast}J^\prime\left(\bar{\lambda}\int_{0}^{T^\ast}f(s,u)du\right)ds+\bar{\lambda}T^\ast\mid
J^\prime(0)\mid}\cdot\sup_{(t,T)\in\mathcal{T}}a(t,T),
\end{align*}
so the boundedness of $f$ follows from Proposition  \ref{prop
ograniczenie a}.

\section{Proofs}\label{Proofs}

This section consists of the proofs of all results presented in
Section \ref{Formulation of the main results}. They appear in the
following order: proofs of Theorems \ref{tw 1 glowne} and
\ref{wniosek z tw1 o eksplozjach }, proof of Theorem \ref{tw 2
glowne} and proofs of Theorems \ref{tw o warunkach koniecznych} and
\ref{tw glowne Tauber}.

\subsection{Proof of Theorem \ref{tw 1 glowne}}
The proofs of Theorem \ref{tw 1 glowne} is preceded by a sequence of
auxiliary lemmas and propositions.

Recall that the sets $\mathcal{T}$ and $\mathcal{T}_{x,y}$, where
$0<x\leq T^\ast$, $0<y\leq T^\ast$ are given by \eqref{dziedzina f}
and \eqref{def zbioru T_x,y}. In the sequel we will use the
notation: $\bar{\mathbb{R}}_{+}:=\mathbb{R}_{+}\cup\{+\infty\}$.

In the following we will consider the function
$h:\mathcal{T}_{x,y}\longrightarrow\bar{\mathbb{R}}_{+}$ given by
\begin{equation}\label{funkcja wybuchowa}
h(t,T):=e^{\varphi(t,T)}, \quad \text{where} \quad \varphi(t,T):=
\left\{\begin{array}{ll} \frac{1}{x-t+y-T}&\text{for}
\ (t,T)\neq(x,y)\\[2ex]
\infty&\text{for} \ (t,T)=(x,y),
\end{array}\right.
\end{equation}
where $0<x<y\leq T^\ast$ and the function
$R_{\alpha,\gamma}:\mathbb{R}_{+}\longrightarrow\mathbb{R}_{+}$
defined as
\begin{gather}\label{funkcja R}
R_{\alpha,\gamma}(z):=\alpha\ln^3\left(\gamma(z+e^2)\right), \qquad
z\geq0, \ \alpha>0, \gamma\geq1.
\end{gather}
It can be verified that
\begin{gather*}
d:=R^{'}_{\alpha,\gamma}(0)=\frac{3\alpha(2+\ln\gamma)^2}{e^2},
\end{gather*}
and that $R_{\alpha,\gamma}$ is concave. Thus

\begin{gather}\label{druga wlasnosc R}
|R_{\alpha,\gamma}(z_1)-R_{\alpha,\gamma}(z_2)|\leq d|z_1-z_2|,
\qquad z_1,z_2\geq0.
\end{gather}
It can also be checked that for a constant $c>0$ s.t.
$\gamma(c\wedge1)\geq1$ we have
\begin{gather}\label{nierownosc dla R(cz)}
R_{\alpha,\gamma}(cz)\geq R_{\alpha,\gamma(c\wedge1)}(z), \qquad
z\geq0.
\end{gather}
Applying Jensen's inequality to the concave function $\ln^3(z+e^2)$
we obtain
\begin{gather}\label{Jensen dla logarytmu}
\ln^3\left(\frac{1}{b-a}\int_{a}^{b}f(x)dx+e^2\right)\geq\frac{1}{b-a}\int_{a}^{b}\ln^3\big(f(x)+e^2\big)dx,
\end{gather}
for any positive integrable function $f$ on the interval $(a,b)$,
$a<b$. \

\begin{prop}\label{prop ciaglosc g}
Let $\alpha>0$, $\gamma\geq1$ be fixed constants such that
$\alpha\gamma>2$ and $\gamma T^\ast>1$. Choose $(x,y)\in\mathcal{T}$
such that $0<x< y<\frac{\alpha}{2}\wedge T^\ast$ and
$\gamma(y-x)>1$. Let the functions $h, R_{\alpha,\gamma}$ be given
by \eqref{funkcja wybuchowa} and \eqref{funkcja R} respectively.
Then the function $g:\mathcal{T}_{x,y}\longrightarrow\mathbb{R}_{+}$
defined by the formula
\begin{equation}\label{mniejsza funkcja}
g(t,T):= \left\{\begin{array}{ll}
e^{-\int_{0}^{t}R_{\alpha,\gamma}\left(\int_{s}^{T}h(s,u)du\right)
ds}\cdot h(t,T)&\emph{for}
\ (t,T)\neq(x,y)\\[2ex]
0&\emph{for} \ (t,T)=(x,y)
\end{array}\right.
\end{equation}
is continuous.
\end{prop}
{\bf Proof:} We need to show continuity of $g$ only in the point
$(x,y)$. Thus consider any point $(t,T)\in\mathcal{T}_{x,y}$ which
is close to $(x,y)$, i.e. s.t. $(t,T)\neq(x,y)$ and $\gamma(T-t)>1$.
Using monotonicity of $R_{\alpha,\gamma}$ and \eqref{Jensen dla
logarytmu} we obtain the following estimation
\begin{align*}
&e^{-\int_{0}^{t}R_{\alpha,\gamma}\left(\int_{s}^{T}h(s,u)du\right)ds}\cdot
h(t,T)=e^{-\alpha\int_{0}^{t}\ln^3\left(\gamma(\int_{s}^{T}h(s,u)du+e^2)\right)ds}\cdot
h(t,T)\\[2ex]
&\leq
e^{-\alpha\int_{0}^{t}\ln^3\left(\gamma\int_{s}^{T}h(s,u)du+e^2\right)ds}\cdot
h(t,T)\leq
e^{-\alpha\int_{0}^{t}\ln^3\left(\frac{\gamma(T-t)}{T-s}\int_{s}^{T}h(s,u)du+e^2\right)ds}\cdot
h(t,T)\\[2ex]
&\leq
e^{-\alpha\int_{0}^{t}\ln^3\left(\frac{1}{T-s}\int_{s}^{T}h(s,u)du+e^2\right)ds}\cdot
h(t,T)\leq
e^{-\alpha\int_{0}^{t}\frac{1}{T-s}\int_{s}^{T}\ln^3(h(s,u)+e^2)duds}\cdot
h(t,T)\\[2ex]
&\leq
e^{-\frac{\alpha}{T}\int_{0}^{t}\int_{s}^{T}\ln^3h(s,u)duds}\cdot
h(t,T)=e^{-\frac{\alpha}{T}\int_{0}^{t}\int_{s}^{T}\varphi^3(s,u)duds+\varphi(t,T)}.
\end{align*}
One can check that
\begin{gather*}
\int_{0}^{t}\int_{s}^{T}\frac{1}{(x-s+y-u)^3}\ du \
ds=\frac{t}{2}\cdot\frac{-T^2-Tt-ty+2Ty+2Tx-tx}{(x-t+y-T)(x+y-2t)(x+y-T)(x+y)},
\end{gather*}
and thus
\begin{gather*}
-\frac{\alpha}{T}\int_{0}^{t}\int_{s}^{T}\varphi^3(s,u)duds+\varphi(t,T)=\left(1-\frac{\alpha
t\left(-T^2-Tt-ty+2Ty+2Tx-tx\right)}{2T(x+y-2t)(x+y-T)(x+y)}\right)\varphi(t,T).
\end{gather*}
Passing to the limit we obtain
\begin{gather*}
\lim_{t\rightarrow x, T\rightarrow
y}(-T^2-Tt-ty+2Ty+2Tx-tx)=y^2-x^2,\\[1ex]
\lim_{t\rightarrow x, T\rightarrow y}(x+y-2t)=y-x, \quad
\lim_{t\rightarrow x, T\rightarrow y}(x+y-T)=x.
\end{gather*}
Hence
\begin{gather*}
\lim_{t\rightarrow x, T\rightarrow y} \left(1-\frac{\alpha
t\left(-T^2-Tt-ty+2Ty+2Tx-tx\right)}{2T(x-t+y-T)(x+y-2t)(x+y-T)(x+y)}\right)=1-\frac{\alpha}{2y}<0,
\end{gather*}
and consequently $\lim_{t\rightarrow x, T\rightarrow
y}g(t,T)=0$.\hfill $\square$

\begin{rem}\label{prop o zwiazku g z h}
The functions $h,R_{\alpha,\gamma},g$ in Proposition \ref{prop
ciaglosc g} satisfy the equation
\begin{gather*}
h(t,T)=e^{\int_{0}^{t}R_{\alpha,\gamma}\left(\int_{s}^{T}h(s,u)du\right)ds}\cdot
g(t,T),\qquad \forall (t,T)\in\mathcal{T}_{x,y}.
\end{gather*}
\end{rem}

\begin{lem}\label{prop Gronwall}
Let $0<t_0\leq T_0<\infty$ and define a set
\begin{gather*}
A:=\Big\{(t,T): t\leq T, \ 0\leq t\leq t_0, \ t\leq T\leq T_0\Big\}.
\end{gather*}
If $d:A\longrightarrow\mathbb{R}_{+}$ is a bounded function
satisfying
\begin{gather}\label{rownanie w Gronwallu}
d(t,T)\leq K \int_{0}^{t}\int_{s}^{T}d(s,u)duds \qquad\forall
(t,T)\in A
\end{gather}
where $0<K<\infty$ then $d(t,T)\equiv0$ on $A$.
\end{lem}
{\bf Proof:} Assume that $d$ is bounded by a constant $M>0$ on $A$.
We show inductively that
\begin{gather}\label{nierownosc indukcyjna}
d(t,T)\leq MK^{n}\frac{(tT)^n}{(n!)^2}, \qquad\forall(t,T)\in A.
\end{gather}
The formula $\eqref{nierownosc indukcyjna}$ is valid for $n=0$.
Assume that it is true for some $n$ and show that it is true for
$n+1$. We have the following estimation
\begin{align*}
d(t,T)&\leq K\int_{0}^{t}\int_{s}^{T}MK^{n}\frac{(su)^n}{(n!)^2}duds=MK^{n+1}\frac{1}{(n!)^2}\int_{0}^{t}s^n(\int_{s}^{T}u^ndu)ds\\[2ex]
&=
MK^{n+1}\frac{1}{(n!)^2}\int_{0}^{t}s^n\left(\frac{T^{n+1}-s^{n+1}}{n+1}\right)ds\leq
MK^{n+1}\frac{1}{(n!)^2}\int_{0}^{t}s^n\frac{T^{n+1}}{n+1}ds\\[2ex]
&=MK^{n+1}\frac{1}{(n!)^2}\frac{t^{n+1}}{(n+1)}\frac{T^{n+1}}{(n+1)}=MK^{n+1}\frac{(tT)^{n+1}}{((n+1)!)^2}.
\end{align*}
Letting $n\longrightarrow\infty$ in $\eqref{nierownosc indukcyjna}$
we see that $d(t,T)=0$. \hfill$\square$

\begin{prop}\label{prop jedynosc rozwiazania h}
Fix $\alpha>0$, $\gamma\geq1$ s.t. $\alpha\gamma>2$ and $\gamma
T^\ast>1$. Let $0<x<y<\frac{\alpha}{2}\wedge T^\ast$,
$\gamma(y-x)>1$, $0<\delta<y$ and
$g:\mathcal{T}_{x,y-\delta}\longrightarrow\mathbb{R}_{+}$ be a
bounded function. Assume that there exists a bounded function
$h:\mathcal{T}_{x,y-\delta}\longrightarrow\mathbb{R}_{+}$ which
solves the following equation
\begin{gather}\label{rownanie w prop. o jedynosci}
h(t,T)=e^{\int_{0}^{t}R_{\alpha,\gamma}\left(\int_{s}^{T}h(s,u)du\right)
ds}\cdot g(t,T),\qquad \forall (t,T)\in\mathcal{T}_{x,y-\delta},
\end{gather}
where $R_{\alpha,\gamma}$ is given by \eqref{funkcja R}. Then $h$ is
uniquely determined in the class of bounded functions on
$\mathcal{T}_{x,y-\delta}$.
\end{prop}
{\bf Proof:} Assume that
$h_1,h_2:\mathcal{T}_{x,y-\delta}\longrightarrow\mathbb{R}_{+}$ are
bounded solutions of \eqref{rownanie w prop. o jedynosci}. Then the
function $\mid h_1-h_2\mid$ is bounded and satisfies
\begin{gather*}
\mid h_1(t,T)-h_2(t,T)\mid\leq \parallel g\parallel\cdot\mid
e^{\int_{0}^{t}R_{\alpha,\gamma}\left(\int_{s}^{T}h_1(s,u)du\right)
ds}-e^{\int_{0}^{t}R_{\alpha,\gamma}\left(\int_{s}^{T}h_2(s,u)du\right)
ds}\mid, \quad \forall (t,T)\in\mathcal{T}_{x,y-\delta},
\end{gather*}
where
\begin{gather*}
\parallel g\parallel=\sup_{(t,T)\in\mathcal{T}_{x,y-\delta}}\mid g(t,T)\mid.
\end{gather*}
As a consequence of the inequality $\mid
e^{x}-e^{y}\mid\leq\max\{e^x,e^y\}\mid x-y\mid$ for
$x,y\in\mathbb{R}$ we have
\begin{gather*}
\mid h_1(t,T)-h_2(t,T)\mid\leq K \int_{0}^{t}\left|
R_{\alpha,\gamma}\left(\int_{s}^{T}
h_1(s,u)du\right)-R_{\alpha,\gamma}\left(\int_{s}^{T}h_2(s,u)du\right)\right|
ds,\quad \forall (t,T)\in\mathcal{T}_{x,y-\delta},
\end{gather*}
where
\begin{gather*}
K:=\parallel g\parallel
\sup_{(t,T)\in\mathcal{T}_{x,y-\delta}}\max_{i=1,2}\Big\{e^{\int_{0}^{t}R_{\alpha,\gamma}\left(\int_{s}^{T}h_i(s,u)du\right)
ds}\Big\}<\infty.
\end{gather*}
In virtue of \eqref{druga wlasnosc R} we have
\begin{gather*}
\mid h_1(t,T)-h_2(t,T)\mid\leq d K \int_{0}^{t}\int_{s}^{T}\mid
h_1(s,u)-h_2(s,u)\mid duds,\qquad \forall
(t,T)\in\mathcal{T}_{x,y-\delta}.
\end{gather*}
In view of Lemma \ref{prop Gronwall}, with $t_0=\min\{x,y-\delta\}$,
$T_{0}=y-\delta$, we have $h_1(t,T)=h_2(t,T)$ for all
$(t,T)\in\mathcal{T}_{x,y-\delta}$. \phantom{a}\hfill$\square$

\begin{prop}\label{prop f>h}
Fix $\alpha>0$, $\gamma\geq1$ s.t. $\alpha\gamma>2$, $\gamma
T^\ast>1$ and the function $R_{\alpha,\gamma}$ given by
\eqref{funkcja R}. Choose $(x,y)$ s.t. $0<x<y<\frac{\alpha}{2}\wedge
T^\ast$, $\gamma(y-x)>1$ and $\delta$ s.t. $0<\delta<y$. Let
$f_{1}:\mathcal{T}_{x,y-\delta}\longrightarrow\mathbb{R}_{+}$, where
 be a bounded function
satisfying inequality
\begin{gather}\label{nierwnosc dla f}
f_1(t,T)\geq
e^{\int_{0}^{t}R_{\alpha,\gamma}\left(\int_{s}^{T}f_1(s,u)du\right)
ds}\cdot g_1(t,T), \qquad \forall (t,T)\in\mathcal{T}_{x,y-\delta},
\end{gather}
where $g_1:\mathcal{T}_{x,y-\delta}\longrightarrow\mathbb{R}_{+}$.
Let $f_2:\mathcal{T}_{x,y-\delta}\longrightarrow\mathbb{R}_{+}$ be a
bounded function solving equation
\begin{gather}
f_2(t,T)=
e^{\int_{0}^{t}R_{\alpha,\gamma}\left(\int_{s}^{T}f_2(s,u)du\right)
ds}\cdot g_2(t,T), \qquad \forall (t,T)\in\mathcal{T}_{x,y-\delta},
\end{gather}
where $g_2:\mathcal{T}_{x,y-\delta}\longrightarrow\mathbb{R}_{+}$ is
a bounded function. Moreover, assume that
\begin{gather}\label{g_1>g_2}
g_1(t,T)\geq g_2(t,T)\geq 0, \qquad
\forall(t,T)\in\mathcal{T}_{x,y-\delta}.
\end{gather}
Then $f_1(t,T)\geq f_2(t,T)$ for all
$(t,T)\in\mathcal{T}_{x,y-\delta}$.
\end{prop}
{\bf Proof:} Let us define the operator $\mathcal{K}$ acting on
bounded functions on $\mathcal{T}_{x,y-\delta}$ by
\begin{gather}\label{def operatora K}
\mathcal{K}k(t,T):=e^{\int_{0}^{t}R_{\alpha,\gamma}\left(\int_{s}^{T}k(s,u)du\right)
ds}\cdot g_2(t,T), \qquad (t,T)\in\mathcal{T}_{x,y-\delta}.
\end{gather}
Let us notice that in view of \eqref{nierwnosc dla
f},\eqref{g_1>g_2} and \eqref{def operatora K} we have
\begin{gather}\label{Kf<f}
\mathcal{K}f_1(t,T)\leq
e^{\int_{0}^{t}R_{\alpha,\gamma}\left(\int_{s}^{T}f_1(s,u)du\right)
ds}\cdot g_1(t,T)\leq f_1(t,T),
\qquad\forall(t,T)\in\mathcal{T}_{x,y-\delta}.
\end{gather}
It is clear that the operator $\mathcal{K}$ is order-preserving,
i.e.
\begin{gather}\label{monotonicznosc K}
k_1(t,T)\leq k_2(t,T)\quad\forall(t,T)\in\mathcal{T}_{x,y-\delta}
\quad\Longrightarrow\quad
\mathcal{K}k_1(t,T)\leq\mathcal{K}k_2(t,T)\quad\forall(t,T)\in\mathcal{T}_{x,y-\delta}.
\end{gather}
Let us consider the sequence of functions:
$f_1,\mathcal{K}f_1,\mathcal{K}^2f_1$,... . In virtue of
\eqref{Kf<f} and \eqref{monotonicznosc K} we see that
$f_1\geq\mathcal{K}f_1\geq\mathcal{K}^2f_1\geq$... .Thus this
sequence is pointwise convergent to some function $\bar{f}$ and it
is bounded by $f_1$, so applying the dominated convergence theorem
in the formula
\begin{gather*}
\mathcal{K}^{n+1}f_1(t,T)=e^{\int_{0}^{t}R_{\alpha,\gamma}\left(\int_{s}^{T}\mathcal{K}^nf_1(s,u)du\right)
ds}\cdot g_2(t,T),\qquad \forall(t,T)\in\mathcal{T}_{x,y-\delta}
\end{gather*}
we obtain
\begin{gather*}
\bar{f}(t,T)=e^{\int_{0}^{t}R_{\alpha,\gamma}\left(\int_{s}^{T}\bar{f}(s,u)du\right)
ds}\cdot g_2(t,T),\qquad \forall(t,T)\in\mathcal{T}_{x,y-\delta}.
\end{gather*}
Moreover, $\bar{f}$ is bounded and thus, in view of Proposition
\ref{prop jedynosc rozwiazania h}, we have $\bar{f}=f_2$. As a
consequence $f_1\geq f_2$ on $\mathcal{T}_{x,y-\delta}$. \hfill
$\square$\\

\noindent
{\bf Proof of Theorem \ref{tw 1 glowne}}\\
Let us notice that for $0<\tilde{\alpha}<a$ and any
$\tilde{\gamma}\geq1$ we have
\begin{gather*}
\lim_{z\rightarrow+\infty}\frac{\tilde{\alpha}\ln^3\left(\tilde{\gamma}(z+e^2)\right)}{a\ln^3z}=\frac{\tilde{\alpha}}{a}<1.
\end{gather*}
Thus \eqref{ogr na Jprim} implies that for $0<\tilde{\alpha}<a$ and
any $\tilde{\gamma}\geq1$ there exists $\tilde{\beta}\in\mathbb{R}$
such that
\begin{gather}\label{szacowanie Jprim przez R_alpha,gamma}
J^\prime(z)\geq
\tilde{\alpha}\ln^3\left(\tilde{\gamma}(z+e^2)\right)+\tilde{\beta}=R_{\tilde{\alpha},\tilde{\gamma}}(z)+\tilde{\beta},
\qquad z\geq0.
\end{gather}
Now fix parameters $\tilde{\alpha}, \tilde{\gamma}$ so that they
satisfy
\begin{gather*}
0<\tilde{\alpha}<a,\quad \tilde{\gamma}\geq1,  \quad
\tilde{\gamma}(\underline{\lambda}\wedge1)\geq1,\quad
\underline{\lambda}\tilde{\alpha}\tilde{\gamma}(\underline{\lambda}\wedge
1)>2, \quad \tilde{\gamma}(\underline{\lambda}\wedge 1) T^\ast>1,
\end{gather*}
and $\tilde{\beta}$ s.t. \eqref{szacowanie Jprim przez
R_alpha,gamma} holds.

Assume that there exists a bounded solution of \eqref{basic3}. Using
\eqref{szacowanie Jprim przez R_alpha,gamma}, \eqref{ograniczenia
lambdy} and \eqref{nierownosc dla R(cz)} the forward rate $f$
satisfies the following inequality
\begin{align}\label{nierownosc dla tilde f }\nonumber
f(t,T)&=e^{\int_{0}^{t}J^\prime(\int_{s}^{T} \lambda(s,u)f
(s,u)du)\lambda(s,T)ds} a(t,T)
\geq e^{\int_{0}^{t}R_{\tilde{\alpha},\tilde{\gamma}}\left(\int_{s}^{T}\lambda(s,u)f (s,u)du\right)\lambda(s,T)ds+\tilde{\beta} t} a(t,T)\\[2ex]\nonumber
&\geq
e^{\underline{\lambda}\int_{0}^{t}R_{\tilde{\alpha},\tilde{\gamma}}\left(\underline{\lambda}\int_{s}^{T}f
(s,u)du\right)ds} e^{\tilde{\beta} t}a(t,T) \geq
e^{\int_{0}^{t}R_{\underline{\lambda}\tilde{\alpha},\tilde{\gamma}(\underline{\lambda}\wedge
1)}\left(\int_{s}^{T}f (s,u)du\right)ds} e^{\tilde{\beta} t}a(t,T)\\[2ex]
&=e^{\int_{0}^{t}R_{\alpha,\gamma}\left(\int_{s}^{T}f
(s,u)du\right)ds} e^{\tilde{\beta} t}a(t,T).
\end{align}
The constants above $\alpha:=\underline{\lambda}\tilde{\alpha}$,
$\gamma:=\tilde{\gamma}(\underline{\lambda}\wedge1)$ satisfy
$\alpha>0$, $\gamma\geq1$, $\alpha\gamma>2$, $\gamma T^\ast>1$.
\noindent Choose $(x,y)\in\mathcal{T}$ such that
$0<x<y<\frac{\alpha}{2}\wedge T^\ast$, $\gamma(y-x)>1$ and fix three
deterministic functions
$h:\mathcal{T}_{x,y}\longrightarrow\bar{\mathbb{R}}_{+}$,
$R_{\alpha,\gamma}:\mathbb{R}_{+}\longrightarrow\mathbb{R}_{+}$,
$g:\mathcal{T}_{x,y}\longrightarrow\mathbb{R}_{+}$ given by
\eqref{funkcja wybuchowa}, \eqref{funkcja R} and \eqref{mniejsza
funkcja} respectively. Recall that, due to Remark \ref{prop o
zwiazku g z h}, they satisfy the equation
\begin{gather}\label{h i g}
h(t,T)=e^{\int_{0}^{t}R_{\alpha,\gamma}\left(\int_{s}^{T}h(s,u)du\right)
ds}\cdot g(t,T),\qquad \forall (t,T)\in\mathcal{T}_{x,y}.
\end{gather}
In virtue of Proposition \ref{prop ciaglosc g} the function $g$ is
continuous on $\mathcal{T}_{x,y}$ and thus bounded. It follows from
Proposition \ref{prop ograniczenie a} that if the constant $K$ is
sufficiently large, then with probability arbitrarily close to $1$,
\begin{equation}\label{initial condition}
e^{\tilde{\beta} t} a(t,T) \geq
g(t,T),\qquad\forall(t,T)\in\mathcal{T}_{x,y}.
\end{equation}

Let us fix $0<\delta<y$ and consider inequality \eqref{nierownosc
dla tilde f } and equality \eqref{h i g} on the set
$\mathcal{T}_{x,y-\delta}$. Then the function $h$ is continuous. In
virtue of Proposition \ref{prop f>h} we have
\begin{align*}
f(t,T)\geq
h(t,T)=e^{\frac{1}{(x-t+y-T)}},\qquad\forall(t,T)\in\mathcal{T}_{x,y-\delta}.
\end{align*}
For any sequence $(t_n,T_n)\in\mathcal{T}_{x,y}$ satisfying
$t_n\uparrow x$, $T_n\uparrow y$ define a sequence
$\delta_n:=\frac{y-T_n}{2}$. Then
\begin{gather*}
f(t,T)\geq
e^{\frac{1}{(x-t+y-T)}},\qquad\forall(t,T)\in\mathcal{T}_{x,y-\delta_n},
\end{gather*}
and consequently $\lim_{n\rightarrow\infty}f(t_n,T_n)=+\infty$ what
is a contradiction with the assumption that $f$ is
bounded.~\hfill$\square$

\subsection{Proof of Theorem \ref{wniosek z tw1 o eksplozjach }}
From the fact that $f$ is a locally bounded solution on
$\mathcal{T}_{x,y}$ we have, as in the  proof of Theorem \ref{tw 1
glowne}, that
\begin{gather*}
f(t,T)\geq h(t,T), \qquad\forall(t,T)\in\mathcal{T}_{x,y-\delta},
\end{gather*}
for each $0<\delta<y$. As a consequence
\begin{gather*}
\lim_{\mathcal{T}_{x,y}\ni(t,T)\rightarrow(T^\ast,T^\ast)}f(t,T)=+\infty.
\end{gather*}
\hfill $\square$

\subsection{Proof of Theorem \ref{tw 2 glowne}}
The proof of Theorem \ref{tw 2 glowne} is preceded by the auxiliary
result.

We can write \eqref{basic3} in the form $f=\mathcal{A}f$, where
\begin{gather}\label{def operatora A}
\mathcal{A}h(t,T):=a(t,T)\cdot
e^{\int_{0}^{t}J^\prime\big(\int_{s}^{T}\lambda (s,u)
h(s,u)du\big)\lambda (s,T) ds},\qquad (t,T)\in \mathcal{T}.
\end{gather}
The proof of Theorem \ref{tw 2 glowne} is based on the properties of
the operator $\mathcal{A}$. If we fix $\omega\in\Omega$ then we can
treat $\mathcal{A}$ as a purely deterministic transformation with
the function $a$ positive and bounded.
\begin{prop}\label{prop o stalej c dla operatora A}
Assume that the function $J^\prime$ satisfies  \eqref{ujemne Jprim}.
Then there exists a positive constant $c$ such that if
\begin{gather*}
h(t,T)\leq c, \quad \forall (t,T)\in\mathcal{T},
\end{gather*}
 for a non-negative function $h$, then
\begin{gather}\label{warunek ograniczajacy na Ah}
\mathcal{A}h(t,T)\leq c, \quad \forall (t,T)\in\mathcal{T}.
\end{gather}
\end{prop}
{\bf Proof:} Let us assume that $h(t,T)\leq c$ for all
$(t,T)\in\mathcal{T}$ for some positive c.  Using the fact that
$J^\prime$ is increasing and $\lambda$ positive, we have
\begin{equation}\label{wyjsciowe szacowanie Ah}\nonumber
\mathcal{A}h(t,T) \leq a(t,T)\cdot e^{J^\prime(\bar {\lambda} c
T^{\ast})  \int_{0}^{t} \lambda (s,T)ds}
\end{equation}
By Proposition \ref{prop ograniczenie a} $a(\cdot,\cdot)$ is bounded
by a positive constant $K=K(\omega)$ and we arrive at the following
inequality
$$
\mathcal{A}h(t,T) \leq K e^{J^\prime(\bar {\lambda} c T^\ast)
\int_{0}^{t} \lambda (s,T)ds },\,\,\,(t,T)\in\mathcal{T}.
$$
It is therefore enough to find a positive constant $c$ such that
\begin{gather*}
\ln K + J^\prime(\bar {\lambda} c T^\ast)\cdot \int_{0}^{t} \lambda
(s,T)ds \leq \ln c,\,\,(t,T)\in\mathcal{T}.
\end{gather*}
If the function $J^\prime$ is negative on $[0, +\infty)$ then it is
enough to take $c=K$. If $J^\prime$ takes positive values then it is
enough to find a positive an arbitrarily large constant $c$ such
that
\begin{gather*}
\ln K + \bar {\lambda} T^\ast \cdot J^\prime(\bar {\lambda} c
T^\ast)\leq \ln c,\,\,(t,T)\in\mathcal{T}.
\end{gather*}
Existence of such $c$ is an immediate  consequence of the assumption
(\ref{ujemne Jprim}).\hfill$\square$\\

\noindent
{\bf Proof of Theorem \ref{tw 2 glowne}}\\
\noindent $i)$ The operator $\mathcal{A}$ is order-preserving, i.e.
\begin{gather*}
h_1\leq h_2 \quad\Longrightarrow\quad \mathcal{A}h_1\leq
\mathcal{A}h_2.
\end{gather*}
The sequence $h_0\equiv 0, \ h_{n+1}:=\mathcal{A}h_n$ is thus
monotonically increasing to $\bar{h}$ and by the monotone
convergence theorem we have
\begin{gather*}
\bar{h}(t,T)=\mathcal{A}\bar{h}(t,T), \qquad
\forall(t,T)\in\mathcal{T}.
\end{gather*}
Moreover, since $h_0\leq c$, where $c=c(\omega)$ is given by
Proposition \ref{prop o stalej c dla operatora A}, $\bar{h}$ is
bounded and thus \eqref{3war na rozw} is satisfied. Moreover, by the
boundedness of $\bar{h}$ it follows that the process
\begin{gather}\label{pomocniczy}
\int_{0}^{t}J^\prime\big(\int_{s}^{T}\lambda (s,u)
\bar{h}(s,u)du\big)\lambda (s,T) ds
\end{gather}
is continuous wrt. $(t,T)\in\mathcal{T}$ for fixed $\omega$. It is
also adapted wrt. $t$. If we replace $\bar{h}(s,u)$ in the above
formula by any bounded field $k(s,u)$ which is adapted wrt. $s$ then
\eqref{pomocniczy} becomes adapted wrt. $t$. As a consequence,
$\bar{h}(\cdot,T)$ is adapted as a limit of the adapted sequence
$\{h_n(\cdot,T)\}$. In virtue of Proposition \ref{prop ograniczenie
a}, the field $\bar{h}$ satisfies \eqref{1war na rozw} and
\eqref{2war na rozw}.

\noindent $ii)$ The function $J'$ is Lipschitz on $[0, +\infty)$ and
therefore we can repeat all arguments from the proof of Proposition
\ref{prop jedynosc rozwiazania h} and the result follows.
\hfill$\square$

\subsection{Proof of Theorem \ref{tw o warunkach koniecznych}}
If $q>0$ then $J^\prime$ satisfies
\begin{gather*}
J^\prime(z)\geq -a+qz+J^\prime_3(0), \qquad z\geq0.
\end{gather*}
If $\nu\{(-\frac{1}{\bar{\lambda}},0)\}>0$ then
\begin{gather*}
J^\prime(z)\geq -a+J_1^{\prime}(z)+J^\prime_3(0), \qquad z\geq0,
\end{gather*}
and due to the formula
\begin{gather*}
J_1^{\prime\prime\prime}(z)=-\int_{-1/{\bar
{\lambda}}}^{0}y^3e^{-zy} \ \nu(dy)\geq 0,
\end{gather*}
$J^\prime_{1}$ is convex and as such it satisfies the inequality
\begin{gather*}
J^\prime_{1}(z)\geq J^{\prime\prime}_{1}(0)z+J^\prime_{1}(0).
\end{gather*}
In both cases \eqref{ogr na Jprim} holds and thus the assertion
follows from Theorem \ref{tw 1 glowne}. \hfill$\square$

\subsection{Proof of Theorem \ref{tw glowne Tauber}}
The following proposition will be useful in the proof of Theorem
\ref{tw glowne Tauber}.

\begin{prop}\label{prop o równowa¿noci}
The following conditions are equivalent
\begin{enumerate}[i)]
\item $\int_{0}^{1}y\nu(dy)=+\infty$,
\item $\lim_{z\rightarrow\infty}J^\prime_{2}(z)=+\infty$.
\end{enumerate}
Moreover, if the measure $\nu$ has a density and
\begin{gather}\label{zachowanie U w prop. o rownowaznosci}
U_{\nu}(x)\sim x\cdot M(x), \quad\text{as} \ x\rightarrow 0,
\end{gather}
where $M$ is such that $M(x){\rightarrow}c>0$ as $x\rightarrow 0$,
then each of the conditions above is equivalent to\\
iii)$\int_{0}^{1}\frac{M(x)}{x}dx=+\infty.$
\end{prop}
{\bf Proof:} Equivalence of $(i)$ and $(ii)$ follows directly from
the dominated convergence theorem. We show equivalence of $(i)$ and
$(iii)$. In virtue of \eqref{zachowanie U w prop. o rownowaznosci}
we have
\begin{gather*}
c\int_{0}^{1}\frac{U_{\nu}(x)}{x^2}dx\leq\int_{0}^{1}\frac{M(x)}{x}dx\leq
C\int_{0}^{1}\frac{U_{\nu}(x)}{x^2}dx
\end{gather*}
for some constants $0<c<C$. We show that $(i)$ holds iff the last
integral diverges. Integrating by parts yields
\begin{align*}
&\int_{0}^{1}\frac{U_{\nu}(x)}{x^2}dx=\int_{0}^{1}\left(\int_{0}^{x}y^2
g(y)dy\right)\cdot\frac{1}{x^2}dx=\left(-\frac{1}{x}\int_{0}^{x}y^2g(y)dy\right)\bigg{\arrowvert}_{0}^{1}+\int_{0}^{1}yg(y)dy\\[2ex]
&=\lim_{x\rightarrow
0}M(x)-\int_{0}^{1}y^2g(y)dy+\int_{0}^{1}yg(y)dy=c+\int_{0}^{1}y^2\nu(dy)+\int_{0}^{1}y\nu(dy),
\end{align*}
where $g$ is a density of $\nu$.\hfill$\square$\\

\noindent{\bf Proof of Theorem \ref{tw glowne Tauber}}\\
Fix $\rho\in(0,+\infty)$. Let us notice that
\begin{gather*}
J_{2}^{\prime\prime}(z)=\int_{0}^{1}y^2e^{-zy}\nu(dy)=\int_{0}^{1}e^{-zy}\mu(dy)
\end{gather*}
is a Laplace transform of the measure $y^2\nu(dy)$. Thus it follows
from Tauberian theorem, see Theorem 2 p.445 in \cite{Feller}, that
the condition \eqref{warunek na zachowanie U w zerze} is equivalent
to
\begin{gather}\label{granica z Taubera}
\lim_{z\rightarrow\infty}\frac{J_2^{\prime\prime}(z)}{\Gamma(\rho+1)z^{-\rho}\cdot
M(\frac{1}{z})}=1,
\end{gather}
where $\Gamma$ stands for the gamma function.

\noindent $i)$ $\rho>1$ \\
Fix $\varepsilon>0$ such that $\rho-\varepsilon>1$.
Using\eqref{ograniczenie funkcji wolno zmieniajacej} we can find
$z_0>0$ such that $M(\frac{1}{z})<z^{\varepsilon}$ for all $z>z_0$.
In virtue of \eqref{granica z Taubera} for any $z>z_0$ we have the
following estimation
\begin{align*}
J^\prime_2(z)&\leq
J^\prime_2(z_0)+2\int_{z_0}^{z}v^{-\rho}M\left(\frac{1}{v}\right)dv\\[2ex]
&\leq
J^\prime_2(z_0)+2\int_{z_0}^{+\infty}v^{\varepsilon-\rho}dv<+\infty.
\end{align*}
Thus condition \eqref{ujemne Jprim} holds because
\begin{gather*}
J^\prime(z)=-a+J_2^{\prime}(z)+J_3^{\prime}(z), \qquad z\geq0,
\end{gather*}
and $J_2^{\prime}$ is a bounded function. The assertion follows from
Theorem \ref{tw 2 glowne}.

To prove $(ii)$ and $(iii)$ let us notice that in view of
Proposition \ref{prop o równowa¿noci} we have
\begin{gather*}
\lim_{z\rightarrow\infty}J^\prime_{2}(z)=+\infty.
\end{gather*}
As a consequence of \eqref{ograniczenie funkcji wolno zmieniajacej}
in the case when $\rho\in(0,1)$ and the assumption \eqref{drugi
warunek na L} if $\rho=1$, we have
\begin{gather*}
\lim_{z\rightarrow\infty}\int_{a}^{z}v^{-\rho}\cdot
M\left(\frac{1}{v}\right)dv=+\infty, \qquad \rho\in(0,1],
\end{gather*}
for any $a>0$. Thus from d'Hospital formula it follows that for any
$a>0$ we have

\begin{gather*}
\lim_{z\rightarrow\infty}\frac{J_2^{\prime}(z)}{\Gamma(\rho+1)\int_{a}^{z}v^{-\rho}\cdot
M(\frac{1}{v})dv}=\lim_{z\rightarrow\infty}\frac{J_2^{\prime\prime}(z)}{\Gamma(\rho+1)
\ z^{-\rho}\cdot M(\frac{1}{z})}=1, \qquad \rho\in(0,1].
\end{gather*}

\noindent $ii)$ $\rho\in(0,1)$

Fix any $\varepsilon>0$ such that $\rho+\varepsilon<1$. By
\eqref{ograniczenie funkcji wolno zmieniajacej} we can find a
constant $a>0$ such that for any $v>a$ we have
$M(\frac{1}{v})>v^{-\varepsilon}$. Then for $z$ sufficiently large
the following estimation holds
\begin{align*}
J_2^{\prime}(z)&\geq
(1-\varepsilon)\Gamma(\rho+1)\int_{a}^{z}v^{-\rho}M(\frac{1}{v})dv\\[2ex]
&\geq(1-\varepsilon)\Gamma(\rho+1)\int_{a}^{z}v^{-\rho}v^{-\varepsilon}dv\\[2ex]
&=\frac{(1-\varepsilon)\Gamma(\rho+1)}{1-(\rho+\varepsilon)}\left(z^{1-(\rho+\varepsilon)}-a^{1-(\rho+\varepsilon)}\right).
\end{align*}
Consequently, $J^\prime$ satisfies \eqref{ogr na Jprim} and the
assertion follows from Theorem \ref{tw 1 glowne}.

\noindent $i)$ $\rho=1$

Let $c>0$ be any positive constant. Using \eqref{pierwszy warunek na
L} we can fix a constant $a>0$ such that
\begin{gather*}
0<1-2c\max_{v\in[a,\infty]}M\left(\frac{1}{v}\right)<1.
\end{gather*}
For large $z$ satisfying
\begin{gather*}
\frac{J_2^{\prime}(z)}{\int_{a}^{z}\frac{1}{v}\cdot
M(\frac{1}{v})dv}\leq 2,
\end{gather*}
we have the following estimation
\begin{align*}
\ln z-cJ_2^{\prime}(z)&=\ln
z-c\frac{J_2^{\prime}(z)}{\int_{a}^{z}\frac{1}{v}\cdot
M\left(\frac{1}{v}\right)dv}\int_{a}^{z}\frac{1}{v}\cdot M\left(\frac{1}{v}\right)dv\\[2ex]
&\geq \ln z-2c\int_{a}^{z}\frac{1}{v}\cdot M\left(\frac{1}{v}\right)dv\\[2ex]
&\geq \ln z-2c[\ln z -\ln a]\max_{v\in[a,z]}M\left(\frac{1}{v}\right)\\[2ex]
&\geq\left(1-2c
\max_{v\in[a,\infty]}M\left(\frac{1}{v}\right)\right)\ln z+ 2c\ln a
\cdot
\max_{v\in[a,\infty]}M\left(\frac{1}{v}\right)\underset{z\rightarrow\infty}{\longrightarrow}\infty.
\end{align*}
Thus condition \eqref{ujemne Jprim} holds because
\begin{gather*}
J^\prime(z)=-a+J_2^{\prime}(z)+J_3^{\prime}(z), \qquad z\geq0
\end{gather*}
and $J_2^{\prime}$ is a bounded function. The assertion follows from
Theorem \ref{tw 2 glowne}. \hfill$\square$

\end{document}